\documentstyle[epsf,epsfig,preprint,aps,amssymb]{revtex}
\draft \preprint{SNUTP 02/034}
\begin{document}
\title{\Large\bf Inflation with blowing-up 
solution of cosmological constant problem
}
\author{Jihn E.
Kim\footnote{jekim@phyp.snu.ac.kr, and 
jekim@th.physik.uni-bonn.de}}
\address{
Physikalisches Institut, Universit\"{a}t Bonn,
D53115 Bonn, Germany, and \\
School of Physics, Seoul National University, Seoul 151-747,
Korea}\maketitle

\begin{abstract}
The cosmological constant problem is how one chooses, without
fine-tuning, one singular point $\Lambda_{eff}=0$ 
for the 4D cosmological constant. 
We argue that some recently discovered {\it weak self-tuning}
solutions can be viewed as blowing-up this one point
into a band of some parameter. These weak self-tuning solutions
may have a virtue that only de Sitter space solutions are
allowed outside this band, allowing an inflationary period.
We adopt the hybrid inflation
at the brane to exit from this inflationary phase and to
enter into the standard Big Bang cosmology.
\\
\vskip 0.5cm\noindent [Key words: cosmological constant,
self-tuning, inflation, brane]
\end{abstract}


\newcommand{\bea}{\begin{eqnarray}}
\newcommand{\eea}{\end{eqnarray}}
\def\beq{\begin{equation}}
\def\eeq{\end{equation}}

\def\one{\bf 1}
\def\two{\bf 2}
\def\five{\bf 5}
\def\ten{\bf 10}
\def\tenb{\overline{\bf 10}}
\def\fiveb{\overline{\bf 5}}
\def\threeb{{\bf\overline{3}}}
\def\three{{\bf 3}}
\def\fb{{\overline{F}\,}}
\def\hb{{\overline{h}}}
\def\Hb{{\overline{H}\,}}

\def\slash#1{#1\!\!\!\!\!\!/}
\def\hf{\frac12}

\def\A{{\cal A}}
\def\Q{{\cal Q}}

\def\p{\partial}

\newcommand{\debug}{\emph{!!! CHECK !!!}}

\newcommand{\dd}{\mathrm{d}\,}
\newcommand{\Tr}{\mathrm{Tr}}
\newcommand{\drep}[2]{(\mathbf{#1},\mathbf{#2})}

\newpage

\def\de{$(0.003\ {\rm eV})^4$}

\section{introduction}

There are three cosmological constants
or vacuum energies which we
try to understand now. The first and most difficult
problem is understanding the vanishing
cosmological constant at the minimum of the
potential, which is the so-called cosmological constant
problem, $\lq\lq$Why is the cosmological constant
zero at a natural scale of the Planck mass
$M_{Pl}=2.44\times 10^{18}$~GeV?"\cite{veltman,weinberg}.
The second is the needed huge vacuum 
energy at the time of inflation\cite{inflation}. 
The third is the present tiny vacuum energy \de \ observed
by the supernova cosmology project 
\cite{supernova}. Among these, the second one is
understood by particle physics models of appropriate
inflaton potentials. One example is the chaotic
inflation that the vacuum energy at the time
of inflation is huge during 60 or more e-folding
time\cite{chaotic}. The third one is also
understood by particle physics models under the name
of quintessence\cite{models}.

At present, understanding the cosmological constant(c.c.)
problem is the most difficult and unsolved problem.
It is a fine-tuning problem in four-dimensional(4D)
space-time. In 4D, in order to have the flat space
the cosmological constant $\Lambda_{eff}$ must be
exactly zero, otherwise the space is curved. In higher
dimensional space-time, this problem has another twist.
The reason is that we require only the effective 
4D space-time is flat after integrating out the
extra space, even though the bulk is curved
with a warp factor. Thus, with extra spaces there is more
freedom. Indeed, the brane models {\it a la} Randall
and Sundrum(RS) give some hope toward understanding
this problem. 

The recent self-tuning solutions
of the cosmological constant problem are worked out
in the RS type models in five
dimensions(5D)\cite{kachru,nilles,kkl,kkl1}. 
These self-tuning solutions are distinguished
by the {\it weak self-tuning solution} and the {\it
strong self-tuning solution}\cite{kkl2}. 
The weak self-tuning solution requires
just the existence of a 4D flat space solution
consistent with the imposed boundary conditions.
On the other hand, the strong self-tuning solution 
requires that the
{\it 4D flat space solution is the only solution} consistent
with the boundary conditions. So far, there does not
exist a strong self-tuning solution\cite{kachru,nilles}. But, 
there are examples of the weak self-tuning 
solutions: with antisymmetric tensor field $H_{MNPQ}$
\cite{kkl,kkl1,kkl2} in RS-II type models\cite{RS2} 
and with the addition\cite{binetruy} 
of the Gauss-Bonnet term\cite{GB}.

The needed inflation in the brane world scenario is to
obtain a de Sitter space solution at $t=-\infty$ and
transform it into a flat space solution at $t=0$ when the
standard Big Bang cosmology commences. The study of 
time dependence is needed.

With the existing weak self-tuning solution, we studied the 
time dependence of a flat space at $t=-\infty$ to a flat 
space at $t=\infty$ as the brane tension $\Lambda_1$, 
located at $y=0$, changes\cite{kkl1}. 
It will be possible to study
the time dependence of a transition from a de Sitter
space solution at $t=-\infty$ to a flat space solution 
at $t=\infty$, with weak self-tuning solution. 
At present we have not obtained a closed form
de Sitter space solution with the weak self-tuning solution,
and hence cannot study this kind of time dependence explicitly. 
But, it has been possible to study the 
time dependence of a flat space at $t=-\infty$ to a flat 
space at $t=\infty$ since we obtained closed form
flat space solutions\cite{kkl1}. 
Therefore, it seems that it is not a trivial problem
to introduce inflation with the self-tuning solutions.

In this paper, we consider inflation 
with the known closed form weak self-tuning
solution\cite{kkl}.
In this Kim-Kyae-Lee(KKL) model, the
flat solutions with a finite Planck mass is possible in a 
finite range of brane tension $\Lambda_1$,
\footnote{This idea can be used in
any self-tuning model with a band of blowing-up
solutions.} 
\begin{equation}\label{bound}
-\sqrt{-6\Lambda_b}<\Lambda_1<\sqrt{-6\Lambda_b}
\end{equation}
where $\Lambda_b$ is the AdS bulk c.c.,
and we set the 5D fundamental mass $M=1$.
Therefore, one anticipates that the flat solutions are not
possible at $|\Lambda_1|>\sqrt{-6\Lambda_b}$ where however
de Sitter space solutions may be possible. If so,
a sufficient inflation is possible in the de Sitter space.
Then, by letting the brane tension $\Lambda_1$ 
fall in the range of Eq.~(\ref{bound}),
one can terminate the inflation and reaches a flat space.
One such example is the hybrid inflation\cite{hybrid} 
at the brane. 

\vskip 0.3cm
\begin{figure}[bt]
\centering \centerline{\epsfig{file=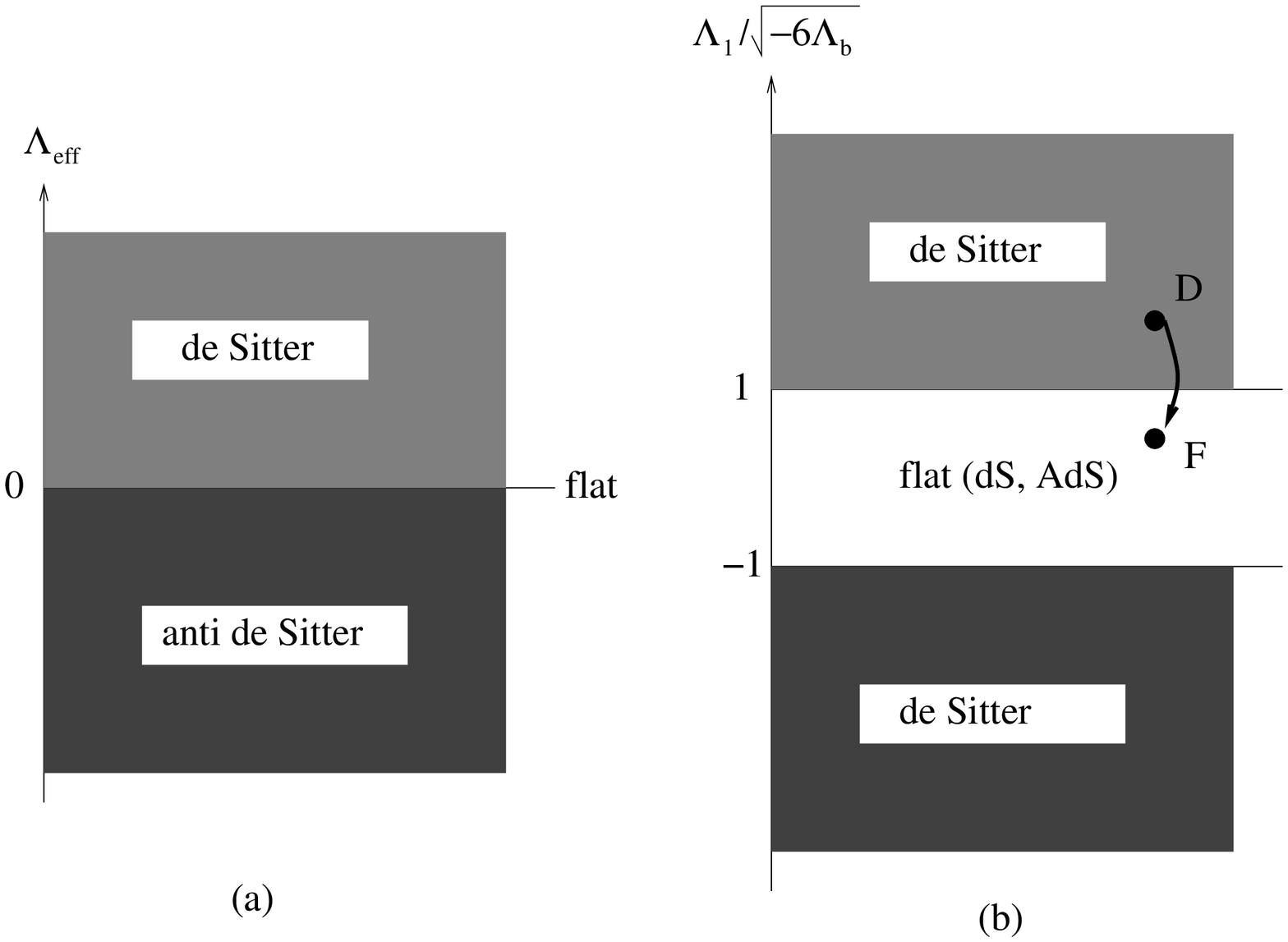,width=110mm}}
\end{figure}
\centerline{ Fig.~1.\ \it  The blowing-up of
the 4D singular point $\Lambda_{eff}=0$ {\rm (a)}}
\centerline{\it into a finite
range of 5D parameter $\Lambda_1$ {\rm (b)}.} 
\vskip 0.3cm

This idea is sketched in Fig.~1.
The cosmological constant problem in 4D is a fine-tuning problem
that the flat space is possible only for $\Lambda_{eff}=0$
as shown in Fig.~1(a). Only at this one point the flat space
is obtained. In our self-tuning solutions\cite{kkl,kkl1,kkl2},
the flat spaces are possible in a finite range of 5D parameters,
as shown in (\ref{bound}). This can be viewed as a blowing-up
of the one point, as shown in Fig.~1(b). One notable feature
is that the KKL model does not allow flat space solutions
outside the range given in (\ref{bound}), namely the inflation
can be naturally introduced in the range of parameters,
$|\Lambda_1|>\sqrt{-6\Lambda}$. Exit from inflation is achieved
by a hybrid inflation at the brane B1, by making $\Lambda_1$
fall in the region allowed in Eq.~(\ref{bound}). In this
case, it is not a fine-tuning since it is possible to have 
a flat space in a finite range of $\Lambda_1$.
The KKL solution, however, is a weak self-tuning solution
since it allows dS and AdS solutions also within the band
(\ref{bound}). If a strong self-tuning solution were
found with the above type of a blowing-up mode, it will
be much more satisfactory.\footnote{Even if the strong 
self-tuning solution of Ref.~\cite{kachru} works
without the problem pointed out in Ref.~\cite{nilles}, 
it does not belong to the class of blowing-up solutions.}

In Sec. II, we review some salient features of the KKL
solution, which is needed for later discussions. 
In Sec. III we show that
there exists the de Sitter space solutions outside the 
region of Eq.~(\ref{bound}).
In Sec. IV, we discuss the hybrid inflation at the brane
so that $\Lambda_1$ falls inside the region given in 
(\ref{bound}) after a sufficient inflation. 
In Sec. V, realization of zero cosmological constant
is discussed. Sec. VI is a brief conclusion.

\section{The weak self-tuning solution: blowing-up model of the
cosmological constant singularity}

There are weak self-tuning solutions in 5D
\cite{kkl,kkl2,binetruy}.
Among these we choose the KKL model\cite{kkl} where a closed form
self-tuning solution was obtained. In this model, we introduce
a three-form field $A_{MNP}\ (M,N,P=0,1,\cdots,4)$, where
its field strength is $H_{MNPQ}$. The relevant action 
is\footnote{In Ref.\cite{kkl2}, another closed form is given with
a logarithmic function of $H^2$.}
\begin{equation}\label{action}
S=\int d^4x \int dy \sqrt{-g}\left(
\frac12 R+\frac{2\cdot 4!}{H^2}-\Lambda_b +{\cal L}_m\delta(y)
\right)
\end{equation}
where $y$ is the fifth coordinate, $H^2=H_{MNPQ}H^{MNPQ}$, 
and we set the fundamental mass
parameter $M=1$ which can be recovered when its explicit
expression is needed. Also, we assume a $Z_2$ symmetry of
the solution, $\beta(-y)=\beta(y)$. The sign of $1/H^2$ was
chosen such that at the vacuum the propagating field $A_{MNP}$
has a standard kinetic energy term.

\subsection{Flat space solutions}

The flat space ansatz for the metric is
\begin{equation}\label{ansatz}
ds^2=\beta^2(y)\eta_{\mu\nu}dx^\mu dx^\nu +dy^2
\end{equation}
where $(\eta_{\mu\nu})={\rm diag.}(-1,+1,+1,+1)$. The
Einstein tensors are
\begin{equation}
G_{\mu\nu}=g_{\mu\nu}\left[3\left(\frac{\beta^\prime}{\beta}\right)^2
+3\left(\frac{\beta^{\prime\prime}}{\beta}\right)\right],\ 
G_{55}=6\left(\frac{\beta^\prime}{\beta}\right)^2
\end{equation}
where prime denotes differentiation with respect to $y$.
The energy-momentum tensors are
\begin{equation}
T_{MN}=-g_{MN}\Lambda_b-g_{\mu\nu}\delta^\mu_M\delta^\nu_N 
\Lambda_1\delta(y)+4\cdot 4!\left(
\frac{4}{H^4}H_{MNPQ}H_N~^{PQR}+\frac12 g_{MN}\frac{1}{H^2}
\right).
\end{equation}

The field equation of $H$,
\begin{equation}
\p_M\left(\sqrt{-g}\frac{H^{MNPQ}}{H^4}\right)=0,
\end{equation}
is satisfied with the following $y$ dependence of the
four-form field
\begin{equation}
H_{\mu\nu\rho\sigma}=\sqrt{-g}\frac{\epsilon_{\mu\nu
\rho\sigma}}{n(y)}.
\end{equation}

The brane B1 located at $y=0$ has a $\delta$-function
discontinuity of $\beta^\prime(0)$, which dictates
the following boundary condition at $y=0^+$,
\begin{equation}\label{BCB1}
\frac{\beta^\prime}{\beta}\Big|_{y=0^+}=-\frac{\Lambda_1}{6}.
\end{equation}
Then, the flat solution is given by
\begin{equation}\label{solution}
\beta(|y|)=\left(\frac{k}{a}\right)^{1/4}\frac{1}{\cosh(
4k|y|+c)^{1/4}}
\end{equation}
where 
\begin{equation}
k=\sqrt{-\frac{\Lambda_b}{6}},\ \ a=\sqrt{\frac{1}{6A}}
\end{equation}
with a positive constant $A>0$, and $c$ is an
integration constant to be determined by the boundary condition
(\ref{BCB1}).

The condition (\ref{BCB1}) determines $c$ as
\begin{equation}\label{crange}
c=\tanh^{-1}\left(\frac{\Lambda_1}{\sqrt{-6\Lambda_b}}\right)
\end{equation}
which allows the flat solution only in the interval
\begin{equation}\label{range}
-\sqrt{-6\Lambda_b} < \Lambda_1 < \sqrt{-6\Lambda_b}.
\end{equation}

One important point of the KKL solution summarized above 
is that the {\it flat 4D space is possible only in the band of
$\Lambda_1$} shown in Eq. (\ref{range}). One can envision
this situation in 5D as a blowing-up of a singular point 
$\Lambda_{eff}=0$ of the {\it flat} 4D case, 
as sketched in Fig. 1.

The 4D Planck mass is expressed as
\begin{equation}\label{Planckm}
M^2_{P,eff}=2M^3\sqrt{\frac{k}{a}}\int_0^\infty
dy\frac{1}{\sqrt{\cosh(4ky+c)}}.
\end{equation}
Thus, the order of the 4D Planck mass squared is
$M^3/\sqrt{ka}$.

\subsection{De Sitter and anti de Sitter space solutions}

The action (\ref{action}) allows the de Sitter and anti de Sitter
space solutions. The ansatz for these curved spaces are taken
as
\begin{equation}\label{ansatzcurved}
ds^2=\beta^2(y)\bar g_{\mu\nu}dx^\mu dx^\nu +dy^2
\end{equation}
where
\begin{equation}\label{dS}
\bar g_{\mu\nu}={\rm diag.\ }(-1, 
e^{2\sqrt{\bar\Lambda}t},
e^{2\sqrt{\bar\Lambda}t},
e^{2\sqrt{\bar\Lambda}t}),\ \ 
\mbox{($dS_4$ background, $\bar\Lambda>0$)}
\end{equation}
and
\begin{equation}\label{AdS}
\bar g_{\mu\nu}={\rm diag.\ }( 
-e^{2\sqrt{-\bar\Lambda}t},
e^{2\sqrt{-\bar\Lambda}t},
e^{2\sqrt{-\bar\Lambda}t},1),\ \ 
\mbox{($AdS_4$ background, $\bar\Lambda<0$)}.
\end{equation}

One important equation, relating $\beta(y)$ and $\beta^\prime(y)$,
is
\begin{equation}\label{prime}
\left|\frac{\beta^\prime}{\beta}\right|
=\sqrt{\frac{\bar k^2}{\beta^2}+k^2-a^2\beta^8}
\end{equation}
where the 4D effective curvature $\bar k$ and the bulk curvature
$k$ are
\begin{equation}
\bar k=\sqrt{\frac{\bar\Lambda}{6}},\ \ k=
\sqrt{\frac{-\Lambda_b}{6}}
\end{equation}

It was shown that there exist de Sitter and
anti de Sitter space solutions\cite{kkl1}.
It can be studied by investigating Eq.~(\ref{prime})
in the limit of $\beta\rightarrow 0$ or $\beta^\prime
\rightarrow 0$\cite{kkl2}.

\subsection{Rescaling of $\beta(0)$}

We can rescale $\beta(0)$ for our convenience. But
this rescaling is possible due to our
implicit {\it assumption that the effective 4D
Planck mass is the definition of a scale.} If a
trans-Plankian physics is present, then the defining
scale can be given at a higher-than-Planck scale and the
following discussion on the rescaling of $\beta(0)$ is 
not allowed.

Let us see how the parameters are changed under the 
rescaling, in the de Sitter space for an explicit
presentation. Defining $\tilde \beta$ as
\begin{equation}
\tilde\beta^2=\frac{\beta^2(y)}{\beta^2(0)},\ \ 
\tilde\beta^2(0)=1,
\end{equation}
we can consistently redefine the space-time coordinates,
\begin{equation}
\tilde t=|\beta(0)|t, \ \ \tilde x^i=|\beta(0)|x^i,
\end{equation}
such that the metric (\ref{dS}) takes the following form,
\begin{equation}
ds^2=\tilde\beta^2(y)\left(
-d\tilde t^2+
d\tilde x^i d\tilde x^i 
e^{2 \sqrt{\bar\Lambda}
\frac{\tilde t}{|\beta(0)|}}
\right)+dy^2.
\end{equation}
The curvature in the new coordinate system is
\begin{equation}\label{tildeL}
\tilde\Lambda=\frac{\bar\Lambda}{\beta^2(0)}.
\end{equation}
The 4D Planck mass in the tilded coordinate is given by
\begin{equation}\label{tildeMP}
\tilde M^2_{P,eff}=\frac{M^2_{P,eff}}{\beta^2(0)}
\end{equation}
where
\begin{equation}\label{Planckmd}
M^2_{P,eff}=2M^3\int_0^\infty\ dy\beta^2(y).
\end{equation}
Therefore, the curvature in units of the Planck mass is
the same in the untilded and tilded coordinates,
\begin{equation}
\frac{\bar\Lambda/M^2}{M^2_{P,eff}}=\frac{\tilde\Lambda/M^2}{
\tilde M^2_{P,eff}}.
\end{equation}

Now, the derivative equation (\ref{prime}) becomes,
\begin{equation}\label{tildeprime}
\left|\frac{\tilde\beta^\prime}{\tilde\beta}\right|
=\sqrt{\frac{\tilde k^2}{\tilde\beta^2}+k^2-a^2\beta^8(0)
\tilde\beta^8}
\end{equation}
where
\begin{equation}
\tilde k^2=\frac{\bar k^2}{\beta^2(0)}.
\end{equation}
Thus, the derivative at $0^+$ is given by
\begin{equation}\label{BCtilde}
|\tilde\beta^\prime|_{0^+}=\sqrt{\tilde k^2+k^2-a^2\beta^8(0)}.
\end{equation}

\section{Existence of de-Sitter-space-only region}

In this section, we study time dependent but spatially
homogenious 4D solutions which can be useful for
cosmological studies.

\subsection{The nearby flat and curved space solutions}

To compare the nearby solutions, it is convenient to use
the tilded coordinate where $\tilde\beta(0)$ is defined to be 1.
Let us consider the nearby solutions for the flat, de Sitter(dS),
and anti de Sitter(AdS) spaces. In this case, three solutions will
be almost identical up to the 
point $y=(very\ large)$ so that the flat
solution tail is negligible. The dS
space solution has the horizon(the point where $\tilde\beta
(y_h)=0$) at $y_h\simeq\infty$. 
Also, the AdS space solution
has the first cycle(up to the point where $\tilde\beta^\prime
(y_A)=0$ and $\tilde\beta(y_A)=\epsilon^+$) extending almost to 
$y_A\simeq\infty$. The three cases of these nearby solutions
are depicted in Fig. 2. Since the AdS solution can be 
discussed similarly as for the dS case, we 
compare below the flat and dS solutions only.

\vskip 0.3cm
\begin{figure}[bt]
\centering \centerline{\epsfig{file=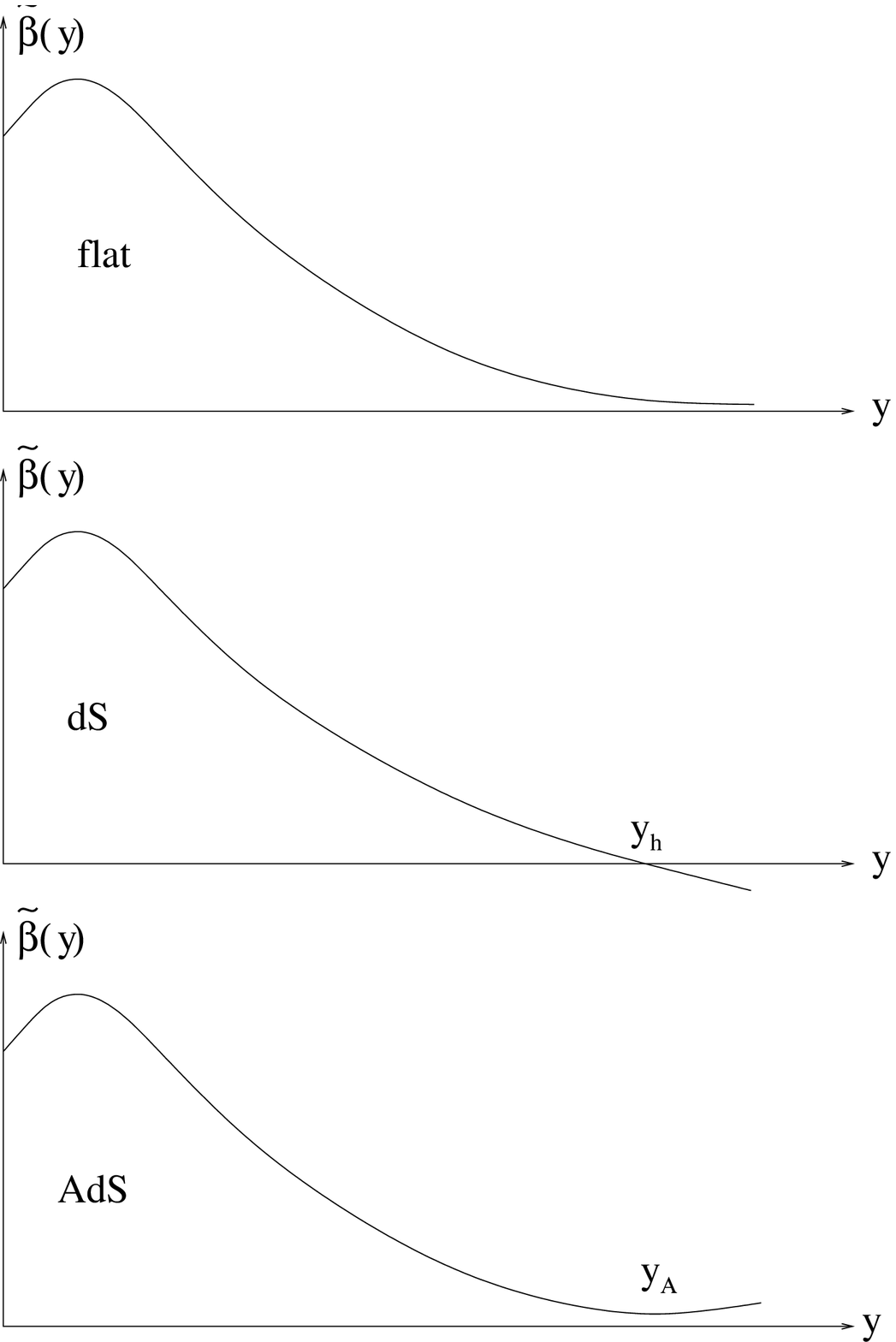,width=110mm}}
\end{figure}
\centerline{ Fig.~2.\ \it  The schematic behaviors of
nearby flat, de Sitter, and anti de Sitter}
\centerline{\it space solutions.
The numerical values of $M_{Pl}$ are almost the same.}
\vskip 0.3cm

In the tilded coordinate, it is easy to compare
the nearby solutions since $\tilde\beta(0)$ is defined
to be 1. The boundary condition at B1, (\ref{BCtilde}),
should be the same for the flat and dS solutions since
we look for solutions with the same $\Lambda_1$. Thus, if both
flat and dS solutions exist,
we have an equality,
\begin{equation}\label{FdS}
\sqrt{k^2-a_{\rm F}^2\beta_{\rm F}^8(0)}
=\sqrt{\epsilon +k^2-a^2_{\rm dS}
\beta^8_{\rm dS}(0)}
\end{equation}
where $\epsilon$ is the infinitesimal curvature $\delta\tilde 
k^2$ near the flat solution, and subscripts F and dS denote
the respective values in the flat and dS space solutions.
We can choose the integration constant $a^2_{\rm dS}$ such that
$a^2_{\rm dS}\beta^8_{\rm dS}(0)=
a^2_{\rm F}\beta^8_{\rm F}(0)+\epsilon$,
satisfying the boundary condition (\ref{FdS}).
Thus, we conclude that there exists a nearby dS space solution
close to the flat space solution, which means that in the band
for the flat solutions (\ref{range}) there exist de Sitter
space solutions. A similar conclusion can be drawn for AdS
solutions. Therefore, in the region where flat space
solutions are allowed, the dS and AdS solutions are 
also allowed. It was shown previously that these dS and AdS 
solutions exist by studying the behavior of $\beta^\prime$
in Ref.~\cite{kkl1}. 

\subsection{De Sitter space solutions at $|\Lambda_1|>\sqrt{
-6\Lambda_b}$}

Let us rewrite (\ref{tildeprime}) as,
\begin{equation}\label{deriv}
|\tilde\beta^\prime|=\sqrt{\tilde k^2+k^2\tilde\beta^2
-\tilde a^2\tilde\beta^{10}}
\end{equation}
where 
\begin{equation}
\tilde a^2=a^2\beta^8(0).
\end{equation}
We know that dS space solutions are possible if
\begin{equation}
\tilde\beta(y_h)=0,\ \ \tilde\beta^\prime(y_h)
\ne 0
\end{equation}
are satisfied\cite{kkl1,kkl2}. Note that the
discontinuity condition of the derivative of
$\beta$ at $y=0$ is
$$
\left|\frac{\tilde\beta^\prime}{\tilde\beta}\right|_{0^+}
=\sqrt{\tilde k^2+k^2-\tilde a^2}=
\left|\frac{\Lambda_1}{6}\right|.
$$
Thus, in the region where flat space solutions are forbidden,
we obtain an inequality,
$$
\sqrt{\tilde k^2+k^2-\tilde a^2}> \sqrt{\frac{-\Lambda_b}{6}}
$$
which is
\begin{equation}
\tilde k^2>\tilde a^2,
\end{equation}
or the effective 4D curvature $\bar\Lambda$ is bounded as
\begin{equation}\label{curvatureB}
\bar\Lambda > 6a^2M^2\beta^{10}(0)
\end{equation}
where we recovered $M^2$. Thus, we conclude that 
in the region $|\Lambda_1|>\sqrt{-6\Lambda_b}$
both flat and AdS solutions are forbidden. {\it Only
dS solutions are allowed with the curvature
bounded from below.}
Note that the de Sitter space solution is
obtained for the negative tension also,
$\Lambda_1<-\sqrt{-6\Lambda_b}$. In this case,
the 4D gravity at the brane is antigravity
and one must be careful in studying cosmological
effects at the negative tension brane\cite{maeda,negative}.

To estimate the order of the
magnitude, let us observe that the flat space solution
(\ref{solution}) shows that $a^2\beta^8(0)$ is proportional
to $k^2$. Thus, we guess that a rough inequality is
$\bar\Lambda>6M^2k^2\beta^2(0)$,
or
\begin{equation}\label{rough}
\bar\Lambda>-\frac{\Lambda_b}{M}\beta^2(0).
\end{equation}

\section{Hybrid inflation at the brane}

Consider the de Sitter space solution where 
$\Lambda_1>\sqrt{-6\Lambda_b}$, i.e. the
positive tension brane 
outside the region given in Eq.~(\ref{range}).
In the previous section, in this region of $\Lambda_1$ 
it was shown that the dS space solution is
the only possibility. Also, the conventional 4D
Einstein equations are obtained at the
positive tension brane\cite{maeda}.
This positive tension 3-brane is called B1. 
Now we can treat $\Lambda_1$ as the vacuum energy at B1.
Namely, $\Lambda_1$ here
is the vacuum energy of the Lagrangian given in
Eq.~(\ref{action}). Therefore, the field evolution at
B1 can be studied as in the usual 4D cosmology. 

Any reasonable inflation model can be adopted to
transform the dS-only region to the region given
in (\ref{range}). Probably, the most modern
and attractive inflationary model is the
{\it hybrid inflation}\cite{hybrid,review} which
we adopt here for the purpose of lucid presentation. 
In the hybrid inflation, at least 
two fields participate in the inflationary 
period. One field $\phi$(the inflaton field) is 
responsible for keeping the potential flat during 
enough e-foldings and another field $\psi$(the
waterfall field) is
responsible for exiting the inflationary period
by radiating light particles. Below we discuss cosmology at
brane B1. 

Since it is sufficient to present our idea 
with the simplest original model of
hybrid inflation\cite{hybrid,linde94}, we adopt this
model in this paper. 
The coupled potential of the inflaton and waterfall fields
is assumed as
\begin{eqnarray}\label{Vhybrid}
&V=\frac12 m^2\phi^2+\frac14 \lambda(\psi^2-\mu^2)^2
+\frac12\lambda^\prime \psi^2\phi^2=V_0 
+\frac12 m^2\phi^2\nonumber\\
&+\frac12(-m_\psi^2
+\lambda^\prime\phi^2)\psi^2+\frac12\lambda^\prime\psi^2
\phi^2
\end{eqnarray}
where $V_0=\frac14 m_\psi^2\mu^2,
m_\psi^2=\lambda \mu^2$, and we treat $\mu^2$
and $m^2$ as large and small parameters, respectively.
Studying the effective mass of the waterfall field 
$\psi$, we can see immediately that $\psi$ sits at
$\psi=0$ if the inflaton field $\phi$ is greater
than the critical 
value $\phi_c$,\footnote{We
will consider the positive $\phi$ region 
during the inflationary period.} 
\begin{equation}
\phi_c\equiv \sqrt{\frac{\lambda}{\lambda^\prime}}~\mu.
\end{equation}
If $\phi$ moves below the critical value, the waterfall
field $\psi$ feels the tachionic potential and rolls down to
the minimum at $\psi=\mu$. It was shown that the time
needed to roll down the hill after $\phi$ passes $\phi_c$
is of order one Hubble time\cite{linde94}.
In the region $|\phi|\gg \phi_c$, the inflaton potential
looks like $V_0+\frac12 m^2\phi^2$, and for a small value
of $m^2$ a sufficient inflation is possible. One can take
$m\sim $~TeV.

During the inflationary period, the height of the potential
is of order
\begin{equation}\label{V0}
V_0=\frac{\lambda}{4}\mu^4.
\end{equation}

During the inflation, we require an almost flat
potential, implying $m^2\ll \frac12\lambda^\prime\mu^2$.
Namely, we require that $\mu^2$ is a large parameter
and $m^2$ is a small parameter.
At the time when $\phi$ reaches $\phi_c$, the Hubble
parameter becomes
\begin{equation}
H^2=\frac{\lambda\mu^4}{12M_{Pl}^2}.
\end{equation}
Requiring $m$ smaller than the Hubble parameter,
$m^2\ll H^2$, we have
\begin{equation}
\mu^2\gg \sqrt{\frac{12}{\lambda}}\ mM_{Pl}.
\end{equation}
With this condition a sufficient inflation is
guaranteed. Furthermore, it was argued that even a
value of $m^2\sim O(H^2)$ does not ruin the needed
inflation\cite{linde94,lythriotto}.

The condition of forbidding the flat space solution,
namely the condition for inflation, is
\begin{equation}\label{inflation}
|\Lambda_1|>\sqrt{6M^3|\Lambda_b|},
\end{equation}
which is equivalent to
\begin{equation}
\mu^4>\sqrt{\frac{96}{\xi\lambda^2}M_{Pl}^2k_a|\Lambda_b|}
\end{equation}
where
\begin{equation}
k_a\equiv \sqrt{ak},
\end{equation}
and $\xi$ is the integral in the definition of
$M^2_{P,eff}=M_{Pl}^2$ in Eqs.~(\ref{Planckm}) and 
(\ref{Planckmd})
\begin{equation}
M^2_{Pl}=\xi\frac{M^3}{k_a}.
\end{equation}
$\xi_{flat}(c)$ of the flat space solution is estimated as
\begin{equation}
\xi_{flat}(c)=\frac12\int_0^\infty dx
\frac{1}{\sqrt{\cosh(x+c)}}\rightarrow\xi_{flat}(0)
=\sqrt{\frac{8}{\pi}}\left(\Gamma(\frac{5}{4})\right)^2
\simeq 1.311.
\end{equation}
If we consider the parameters in the range,
$\xi\simeq 1,\lambda\simeq 1,M=[10^{16}~{\rm GeV},10^{18}~{\rm GeV}],$
then $k_a$ falls in the range $[1.7\times 10^{11}~{\rm GeV},
1.7\times 10^{17}~{\rm GeV}]$. Thus, the inflation condition
becomes
\begin{equation}
\mu>[1.3\times 10^{13}~{\rm GeV}, 1.8\times 10^{18}~{\rm 
GeV}].
\end{equation}
As an eyeball number, let us take
\begin{equation}
-\Lambda_b=10^{14}~{\rm GeV},\ \ k_a=
10^{14}~{\rm GeV},\ \ \xi=1,\ \ \lambda=1. 
\end{equation}
Then
\begin{equation}
M=8.4\times 10^{16}~{\rm GeV},\ \ \mu>2.2\times 10^{15}~
{\rm GeV}.
\end{equation}
From Eq.~(\ref{rough}), we can estimate the allowed
vacuum energy $V_0$ which must be bounded,
\begin{equation}
V_0\sim \bar\Lambda > (1.86\times 10^{13}\ {\rm GeV})^4
\beta^2(0).
\end{equation} 
Since we take $\mu>2.2\times 10^{15}$~GeV (Note that  
$V_0=\frac{\lambda}{4}\mu^4$.), it is consistent
with the above condition for the de Sitter space
curvature $\bar\Lambda$ in the de-Sitter-space-only region.

Thus, there is a reasonable range of parameters satisfying
the inflationary condition.

After a sufficient inflation in the effective 4D cosmology,
the brane tension at B1 changes when the brane field
inflaton $\phi$ becomes smaller than the critical
value $\phi_c$. Within an order of the Hubble time
at $\phi\sim\phi_c$, the brane vacuum energy or the
brane tension $\Lambda_1$ falls in the range of
Eq.~(\ref{range}) which allows flat space solution.
Within this range, however, not only the flat solutions but
also curved solutions are possible. If the dS space solution is
chosen, a further inflation would result. On the other hand,
if a flat space solution is chosen, we exit from inflation
and go into the standard Big Bang cosmology.
The most probable case is that a flat space is chosen
and inflation ends as in the standard hybrid inflation
scenario in the flat space background. The question is
how the flat space solution is chosen after $\Lambda_1$
falls in the range given in Eq.~(\ref{range}).

\section{Vanishing cosmological constant}

In Fig. 1, we sketched the parameter change from
Point D to Point F. Point D allows only the de Sitter
space and inflation is achieved without any difficulty.
However, in the KKL model Point F allows flat, de Sitter
and anti de Sitter space solutions. Only when a flat
space is chosen, we obtain the vanishing cosmological
constant. In this regard, obtaining a strong self-tuning
solution with a blowing-up mode is of utmost importance.
With a strong self-tuning solution, 
then Point F allows only flat space solutions
and we arrive at the zero cosmological constant vacuum
after exiting from the inflationary phase.

With the weak self-tuning solution as discussed in this 
paper, we must rely on an additional principle to choose
flat space solutions. In Ref.~\cite{kkl1}, Hawking's 
probabilistic selection\cite{hawking} was used. 
With the present form of hybrid inflation, this
probabilistic choice can be more elegantly phrased.
The initial condition for the universe is the state
D in Fig. 1(b). With this initial condition,
it is unavoidable to have an
inflationary period. When the inflaton field $\phi$
passes the point $\phi_c$, the parameter set is
given such that the band of Eq.~(\ref{bound})
is chosen. Now there are infinite ways to choose classical
pathes, starting from Point D and ending at Point F.
The probability to end at D with a flat solution is
infinitely larger than the probability to end
with dS or AdS solutions. It is most probable that
flat space solutions are chosen as soon as the path
enters in the band of Eq.~(\ref{bound}). Hawking's
argument\cite{hawking} can be repeated in this 
situation. The Euclidian space integral involves
$\int dy\int d^4x_E(\cdots)$ where $(\cdots)$ represent the
relevant vacuum energy. This integral is maximum
where a flat space is chosen at every point of $y$,
which means that as soon as the bound (\ref{bound})
is satisfied a flat space is chosen and continues to
be so. The flat space obtained in this way 
corresponds to the field values where equations of motion are
satisfied, presumably at the minimum of the potential.
The above argument is notably qualitative. One can question,
$\lq\lq$What is the dominant sourse for ending the inflation?"
We can say qualitatively that
it is the waterfall field in the flat background.
It will be interesting to see more accurately how the inflation
terminates.

Thus, the present tiny vacuum energy above
the point of a true minimum in the
flat background will show up after
the matter energy becomes comparable to dark
energy \de. Eventually, the universe feels the hill
of this tiny tiny potential and will give the vanishing
cosmological constant as soon as this tiny hill
provides an oscillation\footnote{It is stated
in the framework of a very very light pseudo-Goldstone
boson type quintessence\cite{ULGB}.} and 
this oscillating mode behaves like cold dark matter.

\section{Conclusion}

We introduced a reasonable scenario for a
{\it sufficient inflation} and transition
to a vacuum with the {\it vanishing 
cosmological constant} in RS-II type models. 
Here, the observable
sector fields are located at the brane B1.
The necessary ingredient is the de-Sitter-space-only
region in a blowing-up solution of the cosmological
constant problem. 
We discussed the scheme with a previously found
blowing-up solution in the form of
weak self-tuning solution\cite{kkl}. 
This idea can be applied to any blowing-up solution
of the cosmological constant problem with a
de-Sitter-space-only region.
Inflation
is unavoidable for some initial conditions of
the universe due to the existence of de-Sitter-space-only
region. Fine-tuning is avoided because the
flat space solutions are allowed in a finite
range of parameter $\Lambda_1$. 
Exit from inflation can be realized in any reasonable
inflationary model. In particular, we showed that it
is possible to introduce a hybrid inflation,
being consistent
with the blowing-up region of the model we discussed.
Since the example we choose for the blowing-up solution
is a weak self-tuning solution, we must rely on another
principle to choose the flat space after inflation.
For this we adopted Hawking's probabilistic argument.

\acknowledgments
I thank K.-S. Choi, H. B. Kim and H. M. Lee for useful 
discussions. I also thank Humboldt Foundation for the award. 
This work is supported in part by the BK21
program of Ministry of Education, and the KOSEF Sundo Grant.


\begin{thebibliography}{99}

\def\apj#1#2#3{Astrophys.\ J.\ {\bf #1} (#3) #2}
\def\ijmp#1#2#3{Int.\ J.\ Mod.\ Phys.\ {\bf #1} (#3) #2}
\def\mpl#1#2#3{Mod.\ Phys.\ Lett.\ {\bf A#1} (#3) #2 }
\def\nat#1#2#3{Nature\ {\bf #1} (#3) #2}
\def\npb#1#2#3{Nucl.\ Phys.\ {\bf B#1} (#3) #2}
\def\plb#1#2#3{Phys.\ Lett.\ {\bf B#1} (#3) #2}
\def\prd#1#2#3{Phys.\ Rev.\ {\bf D#1} (#3) #2}
\def\prl#1#2#3{Phys.\ Rev.\ Lett.\ {\bf #1} (#3) #2}
\def\prt#1#2#3{Phys.\ Rep.\ {\bf #1} (#3) #2}
\def\sjnp#1#2#3{Sov.\ J.\ Nucl.\ Phys.\ {\bf #1} (#3) #2}
\def\zp#1#2#3{Z.\ Phys.\ {\bf #1} (#3) #2}
\def\jhep#1#2#3{JHEP\ {\bf #1} (#3) #2}
\def\rmp#1#2#3{Rev.\ Mod.\ Phys.\ {\bf #1} (#3) #2}

\bibitem{veltman} M. Veltman, \prl{34}{777}{1975}. 

\bibitem{weinberg} S. Weinberg, \rmp{61}{1}{1989}. For a 
recent summary, see, U. Ellwanger, hep-ph/0203252. 

\bibitem{inflation} A. Guth, \prd{23}{347}{1981};
A. D. Linde, \plb{108}{389}{1982};
A. Albrecht and P. J. Steinhardt, \prl{48}{1220}{1982}.
For a review, see,
A. R. Liddle and D. H. Lyth, \prt{231}{1}{1993} 
[astro-ph/9303019].

\bibitem{supernova} S. Perlmutter {\it et al.}
(Supernova Cosmology Project Collaboration), 
\apj{517}{565}{1999} [astro-ph/9812133].

\bibitem{chaotic} A. D. Linde, \plb{129}{177}{1983}.

\bibitem{models} 
M. Bronstein, {\it Physikalische Zeitschrift 
Sowjet Union} {\bf 3}, 73 (1933); M. $\ddot{\rm O}$zer and M. O. 
Taha, \npb{287}{797}{1987}; B. Ratra and P. J. E. Peebles,
\prd{37}{3406}{1988}; 
C. Wetterich, \npb{302}{645}{1988}; H. Gies and C. Wetterich, 
[hep-ph/0205226];
J. A. Frieman, C. T. Hill and R.
Watkins, \prd{46}{1226}{1992}; 
J. A. Frieman, C. T. Hill, A. Stebins, and I. Waga, 
\prl{75}{2077}{1995} [astro-ph/9505060];
R. Caldwell, R. Dave and P. J. 
Steinhardt, \prl{80}{1582}{1998} [astro-ph/9708069];
J. E. Kim, \jhep{9905}{022}{1999} [hep-ph/9811509]; 
J. E. Kim, \jhep{0006}{016}{2000} [hep-ph/9907528]; 
P. Binetruy, \prd{60}{063502}{1999} [hep-ph/9810553];
C. Kolda and D. H. Lyth, \plb{458}{197}{1999} [hep-ph/9811375];
T. Chiba, \prd{60}{083508}{1999} [gr-qc/9903094];
P. Brax and J. Martin, \plb{468}{40}{1999} [astro-ph/9905040];
A. Masiero, M. Pietroni, and F. Rosati, \prd{61}{023504}{2000}
[hep-ph/9905346];
M. C. Bento and O. Bertolami, {\it Gen. Rel. Grav.} {\bf 31}
(1999) 1461 [gr-qc/9905075];
F. Perrotta, C. Baccigalupi, and S. Matarrese, \prd{61}{023507}{2000}
[astro-ph/9906066];
A. Arbey, J. Lesgourgues, and P. Salati, \prd{65}{083514}{2002}
[astro-ph/0112324]; 
A. Albrecht, C. P. Burgess, F. Ravndal, and C. Skordis, 
\prd{65}{123507}{2002} [astro-ph/0107573]; 
C. T. Hill and A. K.
Leibovich, hep-ph/0205237.

\bibitem{kachru} N. Arkani-Hamed, S. Dimopoulos, N. Kaloper, and
R. Sundrum, \plb{480}{193}{2000} [hep-th/0001197];
S. Kachru, M. B. Schulz, and E. Silverstein, \prd{62}{045021}{2000}
[hep-th/0001206].

\bibitem{nilles}
S. F$\ddot{\rm o}$rste, Z. Lalak, S. Lavignac, and H. P. Nilles,
\plb{481}{360}{2000} [hep-th/0002164]; \jhep{09}{034}{2000}
[hep-th/0006139].

\bibitem{kkl} J. E. Kim, B. Kyae, and H. M. Lee, 
\prl{86}{4223}{2001} [hep-th/0011118].

\bibitem{kkl1} J. E. Kim, B. Kyae, and H. M. Lee,
\npb{613}{306}{2001} [hep-th/0101027]; 
J. Korean Phys. Soc. {\bf
40} (2002) 207 [hep-th/0201055]. 

\bibitem{kkl2} J. E. Kim and H. M. Lee, \jhep{0209}{052}{2002}
[hep-th/0207260].

\bibitem{RS2} L. Randall and R. Sundrum, \prl{83}{4690}{1999}
[hep-th/9906064].

\bibitem{binetruy} P. Binetruy, C. Charmousis,
S. C. Davis, and J. F. Dufaux, \plb{544}{183}{2002} 
[hep-th/0206089];
A. Jakobek, K. A. Meissner, and M. Olechowski, hep-th/0206254.

\bibitem{GB} B. Zwiebach, \plb{156}{315}{1985}; 
J. E. Kim, B. Kyae, and H. M. Lee, \prd{62}{045013}{2000}
[hep-ph/9912344];
\npb{582}{296}{2000} [hep-th/0004005].

\bibitem{hybrid} A. D. Linde, \plb{249}{18}{1990};
F. C. Adams and K. Freese, \prd{43}{353}{1991};
A. D. Linde, \plb{259}{38}{1991}.

\bibitem{maeda} T. Shiromizu, K. Maeda, and M. Sasaki,
\prd{62}{024012}{2000} [gr-qc/9910076].

\bibitem{negative}
C. Csaki, M. Graesser, C. Kolda, and J. Terning,
\plb{462}{34}{1999} [hep-ph/9906513];
J. M. Cline, C. Grojean, and G. Servant, \prl{83}{4245}{1999} 
[hep-ph/9906523]
J. Garriga and T. Tanaka, \prl{84}{2778}{2000} [hep-th/9911055].

\bibitem{review} A. R. Liddle and D. H. Lyth, in {\it
Cosmological Inflation and Large Scale Structure}
(Cambridge University Press, Cambridge, England, 2000),
p. 216.

\bibitem{linde94} A. D. Linde, \prd{49}{748}{1994} 
[astro-ph/9307002].

\bibitem{lythriotto} D. H. Lyth and A. Riotto, \prt{314}{1}{1998}
[hep-ph/9807278].

\bibitem{hawking} S. W. Hawking, \plb{134}{403}{1984}.

\bibitem{ULGB} J. E. Kim and H. P. Nilles,
SNUTP 02/032.

\end{thebibliography}
\end{document}